\documentclass[12pt]{article} 
\usepackage{amsfonts}
\usepackage{amssymb}
\usepackage{cite}

\def\be{\begin{equation}} \def\ee{\end{equation}}
\def\bea{\begin{eqnarray}} \def\eea{\end{eqnarray}} \def\ba{\begin{array}}
\def\ea{\end{array}} \def\ben{\begin{enumerate}} \def\een{\end{enumerate}}
  
\def\lll{\label}
 
\begin{document}
\newcommand{\half}{{\textstyle\frac{1}{2}}}
\newcommand{\eqn}[1]{(\ref{#1})}
\newcommand{\npb}[3]{Nucl. Phys. {\bf B#1} ({#2}) {#3}}
\newcommand{\pr}[3]{Phys. Rep. {\bf#1} ({#2}) {#3}}
\newcommand{\prl}[3]{Phys. Rev. Lett. {\bf#1} ({#2}) {#3}}
\newcommand{\plb}[3]{Phys. Lett. {\bf B#1} ({#2}) {#3}}
\newcommand{\prd}[3]{Phys. Rev. {\bf D#1} ({#2}) {#3}}
\newcommand{\hepth}[1]{hep-th/{#1}}
\newcommand{\grqc}[1]{[gr-qc/{#1}]}
 
\def\a{\alpha}
\def\b{\beta}
\def\g{\gamma}\def\G{\Gamma}
\def\d{\delta}\def\D{\Delta}
\def\ep{\epsilon}
\def\et{\eta}
\def\z{\zeta}
\def\t{\theta}\def\T{\Theta}
\def\l{\lambda}\def\L{\Lambda}
\def\m{\mu}
\def\f{\phi}\def\F{\Phi}
\def\n{\nu}
\def\p{\psi}\def\P{\Psi}
\def\r{\rho}
\def\s{\sigma}\def\S{\Sigma}
\def\ta{\tau}
\def\x{\chi}
\def\o{\omega}\def\O{\Omega}
\def\k{\kappa}
\def\pa {\partial}
\def\ov{\over}
\def\br{\nonumber\\}
\def\ud{\underline}

\begin{flushright}
hep-th/0102110\\
\end{flushright}
\bigskip\bigskip
\begin{center}
{\large\bf 
 Massive Type IIA Theory on K3          
}\footnote{Work 
supported by:
AvH -- the Alexander von Humboldt foundation,
DFG -- the German Science Foundation,
GIF -- the German--Israeli
Foundation for Scientific Research,
DAAD -- the German Academic Exchange Service.}

\vskip1cm
{\sc Michael Haack, Jan Louis and 
Harvendra Singh }\footnote{email:
{\tt
haack, j.louis, singh@physik.uni-halle.de}}
\vskip0.5cm
 {Fachbereich Physik, Martin-Luther-Universit\"at Halle-Wittenberg,\\
Friedemann-Bach-Platz 6, D-06099 Halle, Germany}

\vskip 2cm
\end{center}
\bigskip
\centerline{\bf ABSTRACT}
\bigskip

\begin{quote}

In this paper we study $K3$ compactification
of ten-dimensional massive type IIA theory with all 
possible Ramond-Ramond background fluxes 
turned on. The resulting six-dimensional theory is
a new
massive (gauged) supergravity with an 
action that is manifestly invariant under an 
$O(4,20)/ O(4)\times O(20)$ duality symmetry.
We discover that this six-dimensional theory 
interpolates between
vacua of ten-dimensional massive IIA supergravity
and vacua of massless IIA supergravity
with appropriate background fluxes turned on.
%
This in turn suggests a new 11-dimensional interpretation 
for the massive type IIA
theory. 
\end{quote}
\newpage

\section{Introduction}

Recently there has been renewed interest in the study of 
gauged/massive supergravity theories
due to the AdS/CFT correspondence \cite{maldacena}. 
In gauged supergravity theories 
either a subgroup of the $R$-symmetry group 
(the automorphism group of the supersymmetry algebra) 
or isometries of the scalar manifold
are gauged by 
some of the vector fields in the 
spectrum \cite{gsugra}. 
Such theories can be constructed from their
ungauged `cousins' by adding appropriate terms
to the action or field equations 
and changing the supersymmetry
transformation laws accordingly. 
This procedure does not change the spectrum
and the number of supercharges but generically
does change the properties of the ground state.
For example, a Minkowskian spacetime which always
is a solution of the
ungauged theory  ceases
to be a ground state of the gauged supergravity
in most cases. 
Instead the supersymmetric ground states are
often of the anti-de-Sitter (AdS) type  
or  domain-wall solutions. 

A particular subclass of gauged supergravities
are the massive supergravities. 
In these theories some of the vector 
(or tensor) fields become massive
through a generalized Higgs mechanism.
Such theories occur for a specific gauging 
which allows the possibility of a Higgs-type
mechanism. 
A prominent example 
is the massive type IIA supergravity
in $D=10$ constructed by Romans \cite{roma}.
In string theory massive supergravities
typically arise in lower dimensions through generalized Scherk-Schwarz
reduction \cite{ss}
provided that some  field strength $F_{p+1}$
of the $p$-form tensor field $A_p$ is given a 
non-trivial background value (flux) along the
compact directions \cite{berg, cow,lavpop}. 
Such background fluxes can be turned on
consistently if the action or the field equations
depend on $A_p$ 
only through its field strength $F_{p+1}$. 

Gauged supergravities have also been 
studied in the context of string dualities and 
$D$-branes. 
For example,
the massive type IIA supergravity has a domain wall solution
which  preserves $1/2$ of the
32 supercharges of type IIA \cite{berg} and can be given an
interpretation of
a type IIA D-8-brane \cite{pol}.
This observation has led to the search for possible duality connections
involving massive supergravity theories analogous to the existing
U-duality relations in the massless cases.  
This required the construction of other massive 
supergravities in lower dimensions through generalized dimensional reduction 
\cite{berg,cow,blo,bs,mo,lavpop, maha, lupop, kkm,km,cvetic,hull}.
However, it has remained an interesting and open question 
to what extent these generalized compactifications
 respect the duality
properties of the massless cases.
Along this line, Kaloper and Myers \cite{km} 
showed that a generalized Scherk-Schwarz 
reduction of the heterotic string on a 
$d$-torus can still be written in a manifestly
$O(d,d+16)$ invariant form provided T-duality 
transformations also act on
the background fluxes.

In a parallel line of developments
 Calabi-Yau compactifications with background fluxes 
have also been studied 
because of their
phenomenological properties 
\cite{DSHD,S,PS,BB,JM,LOW,GVW,DRS,G,TV,M2,ST,GSS,HLM,
CKLT,EW,LOW2,LOSW,NOY,BG,CK}. 
One finds that background
fluxes generically generate a potential 
for some of the moduli fields of the 
theory without fluxes and as a consequence the moduli space
-- and hence the arbitrariness of the theory --
is reduced.
In addition the resulting ground states 
can break supersymmetry spontaneously.

In this paper we
study a generalized $K3$ reduction
of ten-dimensional massive type-IIA theory
with all possible background fluxes turned on.
Our goal is to investigate the fate of the
perturbative $O(4,20)$ 
duality symmetry and the non-perturbative
S-duality with the heterotic string
compactified on $T^4$. We also discuss the relation of massive type
IIA theory  with M-theory.
The paper is organized as follows. 
In section~2 we recall the massive type II theory
and derive its  $K3$ compactification  with all 
R-R background fluxes turned on. 
We find that the resulting six-dimensional theory
is a new massive (gauged) supergravity 
with manifest $O(4,20)$ 
symmetry.
In section~3 we study domain wall solutions
of this massive supergravity.
We show that 
the D-8 brane of massive type IIA
when wrapped on $K3$ is T-dual to a 
solution of massless type IIA with an appropriate
four-form flux turned on.
Thus T-duality 
interpolates 
between vacua of massive and massless type IIA theories.
This property also allows to further relate massive type II vacua
to eleven dimensional solutions. This is discussed in section~4
 where massive IIA
theory is given a concrete 
eleven-dimensional interpretation.
Finally we conclude in the section~5.

\section{$K3$ Compactification of Massive Type IIA Supergravity}
The type IIA supergravity in ten dimensions, which describes the low
energy limit of type IIA superstrings,
contains in the massless spectrum the graviton
$\hat g_{MN}$, the dilaton
$\hat\f$, an NS-NS two-form $\hat B_2$, a R-R one-form $\hat A_1$ 
and a R-R three-form $\hat C_{3}$. The fermionic
fields consist of two gravitini and two Majorana spinors.  
It was shown by Romans \cite{roma} that
this supergravity can
be generalized to include a mass term for the $\hat B$-field without 
disturbing the supersymmetry. The $\hat B$-field becomes massive through a
Higgs type mechanism in which it eats the vector field $\hat A_1$. 
The supersymmetric action for massive IIA theory in the string
frame is given by \cite{roma}
\begin{eqnarray} \label{massive2a}
S=\int \bigg[ &&e^{-2\hat\f}({1\ov 4}\hat
R~^{\hat\ast}1+d\hat\f~^{\hat\ast} d\hat\f -
{1\over2}\hat H_3 ~^{\hat\ast}\hat H_3) 
-{1\over2}\hat F_2 ~^{\hat\ast}\hat F_2 -{1\over2}\hat F_4
~^{\hat\ast}\hat F_4- {m^2\ov 2 }~^{\hat\ast} 1 \br
&& + d\hat C_3 d\hat C_3 \hat B_2 + 2d\hat
C_3 d\hat A_1 \hat B_2^2+{4\ov3}d\hat A_1 d\hat A_1\hat B_2^3
+{4\ov3}m d\hat C_3\hat B_2^3+2md\hat A_1 \hat B_2^4+{4\ov5}m^2\hat
B_2^5\bigg]
\br &&+\ {\rm
fermionic~~ part} \ ,
\lll{1a}
\end{eqnarray}
where we have adopted the notation that 
every product of forms is understood 
as a wedge product.
The signature of the metric is $(-+\ldots+ )$
and  for a $p$-form  we use the convention
\be
F_p={1\over
p!}F_{M_1...M_p}dx^{M_1}\wedge...\wedge dx^{M_p}\ , \ee
while the Poincare dual is given by
\bea
&&^{\hat\ast}F_{p}={1\over p!(10-p)!} F_{M_1...M_p}
\epsilon^{M_1...M_p}_{~~~~~~~M_{p+1}...M_{10}}~dx^{M_{p+1}}\wedge...\wedge
dx^{M_{10}} , \br
&&~^{\hat\ast \hat\ast}F_{p}=-(-1)^{p(10-p)}F_p\ ,
\eea
and $~^{\hat\ast} 1 =\sqrt{- \hat g}\, d^{10}x$. 
$m$ is the mass parameter and
the various field strengths in the Lagrangian \eqn{1a}
are defined as 
\be
\hat H_3=d\hat B_2\ ,
\qquad 
\hat F_2= d\hat A_1 +2m\hat B_2\ ,
\qquad
\hat F_4= d\hat C_3
+2\hat B_2d\hat A_1+2m\hat B_2^2\ .
\ee
$\hat A_1$ and $\hat C_3$ only appear through their derivatives 
in the Lagrangian \eqn{1a} and thus obey the standard 
$p$-form gauge invariance $A_p \to A_p + d\Lambda_{p-1}$.
The two-form $ \hat B_2$ on the other hand also appears
without derivatives but nevertheless the `Stueckelberg'
gauge transformation
\be
\delta \hat A_1=-2m\Lambda_1\ ,\qquad
\delta \hat B_2=d\Lambda_1\ ,\qquad 
\delta\hat C_3=-2\Lambda_1 d\hat A_1\ 
\ee
leave the Lagrangian invariant. Finally,
the conventional massless type IIA  theory 
is recovered from the action
(\ref{1a}) in the limit $m\to 0$.  

Before we turn to the compactification let
us first recall some facts about $K3$ manifolds. 
$K3$ is a compact, 
Ricci flat complex manifold with Betti numbers $b_0=1, b_1=0,
b_2=22, b_3=0, b_4=1$. Thus there exist 22 harmonic two-forms
$\O_2^i~(i=1,\ldots,22)$ and
their intersection matrix 
\be
  \eta_{ij}= 
\int_{K3} \O_2^i\wedge\O_2^j
\ee
is a Lorentzian metric with signature $(3,19)$. 
We choose conventions for the two-forms $\Omega_2^i$ 
such that $\eta$ is given by
\be \label{eta}
\eta_{ij}=\left(\ba{ccc} 0&0&\sigma\\ 
0&I_{16}&0\\ \sigma &0&0 \ea\right)\ , \qquad {\rm where} \qquad 
\sigma=\left(\ba{ccc} 0&0&1 \\ 0&1&0 \\ 1&0&0 \ea\right)\ ,
\ee
and $I_n$ represents the $n$-dimensional 
identity matrix.
Since $K3$ is four-dimensional the Hodge-dual of 
the harmonic 2-forms can again be expanded in terms
of two-forms. More precisely one has 
\be\label{omegadual}
^\ast\O_2^i=M^i_{~j}~\O_2^j \ ,
\ee
where the matrix $M^i_{~j}$ depends
on the $3\times 19=57$ moduli parameterizing
the deformations of the metric $\delta g_{mn}$
of constant volume on $K3$.
For $K3$  $~^{\ast\ast}~ = ~1$ holds which implies
\be
M^i_{~j}M^j_{~k} = \delta_k^i\ , \qquad 
\eta_{ij} M^j_{~k} = \eta_{kj} M^j_{~i}\ ,
\ee
so that $M^j_{~i} \eta_{jk} M^k_{~l} = \eta_{il}$.
Thus $M$ 
is an element of the coset
$O(3,19)/ O(3)\times O(19)$  \cite{harvey,DLM}.

Let us now turn to the compactification of the 
massive type IIA theory on $K3$. 
The standard Kaluza-Klein 
reduction considers the theory in a spacetime background 
$M_D\times K_d$, where $M_D$
is a non-compact $D$-dimensional manifold with 
Lorentzian signature
while $K_d$ is a $d$-dimensional compact manifold.
This ansatz is consistent whenever the spacetime
background satisfies the $D+d$-dimensional field equations.
However, for massive type IIA there are no direct product $M_6\times
K3$ solutions; instead the ground states are domain-wall solutions
\cite{berg}. 
A similar situation occurs in
the $S^1$ compactification of massive type IIA discussed in \cite{berg}. 
However, as argued there 
one can also expand around a solution 
which strictly speaking is not
a direct product $M_D\times K_d$, but rather the compact manifold  $K_d$
is allowed to vary over the spacetime manifold $M_D$.\footnote{Or in
other words the moduli of $K_d$ are not constant in 
the background but vary over $M_D$.}
For the case at hand such a solution exists and 
is given by  the D-8-brane solution of massive type IIA theory
wrapped on a $K3$ \cite{berg,BP}. We discuss this solution 
in the next section and will find that it can be interpreted 
as the product of a $6$-dimensional domain wall 
with a warped $K3$ whose volume varies over the transverse direction.
This spacetime dependence of the volume ensures that the equations 
of motion of massive type IIA theory 
are fulfilled.  

Thus for the 10-dimensional metric we take the standard ansatz
\be\label{metric}
\hat G_{MN}\, (x,z)= \left(\ba{cc}g_{\m\n}(x)&0\\
0&g_{mn}^0(z) +\delta g_{mn}(x,z)\ea\right)\ ,
\ee 
where $g_{\m\n}(x)$ is the metric and 
$x$ are the coordinates of $M_6$.
$g_{mn}^0(z)$ is a fixed background metric on $K3$
(with coordinates $z$)
and $\delta g_{mn}(x,z)$ denotes the allowed deformations
of the $K3$ metric.
These deformations are parameterized
by $57+1$ moduli where the extra modulus corresponds to the
overall volume of $K3$.\footnote{The metric on the
moduli space of sigma models on $K3$ does 
not change with respect to the standard Kaluza-Klein 
reduction of massless type IIA theory and can therefore be 
taken from that case \cite{sen,harvey,DLM}. }

For the dilaton and the two-form $\hat B_2$ we take the standard
ansatz precisely as in massless type IIA theories
\be
\hat\f(x,z)=\hat \f(x)\ ,\qquad
\hat B_2(x,z)=B_2(x)+b^i(x)~\O_2^i(z)\ ,
\ee
where $B_2(x)$ is a two-form in $D=6$
and the $b^i(x)$ are 22 additional scalar fields.
(Thus the total number of scalars is $58+22+1=81$.)

For the one-form $\hat A_1$ and the three-form $\hat
C_3$
we take a generalized 
Kaluza-Klein ansatz where background values (fluxes)
of the corresponding field strengths are included 
\bea
&& d\hat A_1(x,z)=dA_1(x)+m^i~\O_2^i(z)\ ,\br
&&d\hat C_3(x,z)=dC_3(x)+dC_1^i(x)~\O^i_2(z) -\tilde m~\O_4(z)\ .
\lll{ansatz}
\eea
This ansatz introduces 23 new mass parameters $m^i$ and
$\tilde m$ parameterizing fluxes along the 22 two-cycles 
and the four-cycle on $K3$. 
This generalization is possible since  
$\hat A_1,\hat C_3$ appear in the action 
 only with derivative couplings (i.e.\ via their
field strength) and an appropriate background value
can be consistently turned on \cite{lavpop}.

Altogether,
the bosonic spectrum of the reduced six-dimensional
theory consists of the graviton $g_{\m\n}$, the two form
$B_{2}$, 22+1 one-form gauge fields $(C^i_1,A_1)$, a three-form
$C_3$ and 81 scalar fields $\phi,b^i,\delta g_{mn}$.
{}From their definition it is clear that
 $C_1^i$ and $b^i$ both transform in the
vector representation of $O(3,19)$. 
In $D=6$ a three-form is Poincare dual to a one-form
and thus the above spectrum combines into a gravitational
multiplet consisting of
the graviton, the two-form, four vector fields and the dilaton
and 
20 vector multiplets each containing a one-form and four scalars.

In order to obtain the action of the massless
modes for this theory we
substitute the ansatz \eqn{metric}-\eqn{ansatz} into the action
\eqn{1a}. The resulting six-dimensional bosonic action reads 
\bea
S_6&=&\int\bigg[{1\ov4}e^{-2\f}\left( R~^\ast 1+4 d\f~ ^\ast d\f -
2 H_3 ~^\ast H_3 +{1\over 8} {\rm Tr}  d{\cal M}^{-1} ~^\ast
d{\cal M}\right)
-{1\over2} {\cal F}_2^a ({\cal M}^{-1})_{ab}~^\ast {\cal F}_2^b 
\br &-&
{1\over2} m^a({\cal M}^{-1})_{ab}~^\ast m^b
+ B_2 {\cal F}_2^a {\cal L}_{ab} {\cal F}_2^b  
- 2 B_2^2 m^a {\cal L}_{ab} {\cal F}_2^b   
+{4\over3} B_2^3 m^a {\cal L}_{ab} m^b \bigg] ,
\lll{action}
\eea
where 
\be\label{deff}
2\f=2\hat\f-\ln \o\ ,\qquad H_3=dB_2\ ,\qquad
{\cal F}^a_2 =d {\cal A}^a_1 + 2m^a B_2\  .
\ee
The index $a$ takes values $a=1,\ldots,24$
and we have defined
\be
m^a\equiv (\tilde m,~m^i,~m)\ ,
\qquad
{\cal A}_1^a\equiv (\tilde A_1,~C_1^i,~A_1)\ .
\ee
$d\tilde A_1$ is the Poincare dual of the 4-form $dC_3$ defined as 
\be\label{dualdef}
d\tilde A_1=-\omega~^\ast
(dC_3+2B_2dA_1+2mB_2^2)+2b_i(dC_1^i+b^idA_1)+2(-\tilde
m+2b_i (m^i+m~b^i))B_2\ ,
\ee
where the indices $i,j$ are contracted with the metric $\eta_{ij}$
and $\omega(x)$ is the modulus associated
with the overall volume $V$ of the $K3$
\be \label{volume}
V = \omega (x) \int_{K3} \Omega_4\ .
\ee
{}From eq.\ (\ref{dualdef}) we learn that
$d\tilde A$ is  an $O(3,19)$ invariant 2-form field strength. 
Finally, the scalar matrix ${\cal M}^{ab}$ which 
appears in the action 
(\ref{action}) depends on 
the $58$ moduli of $K3$ and the 22 $b^i$
in the following way
\be
{\cal M}^{-1}= {\cal V}^T{\cal V}\ ,\qquad
{\cal
V}=\left(\begin{array}{ccc}\o^{-\frac12}\ \ &-2\o^{-\frac12} \bar b\ \
&-2\o^{-\frac12}\bar b 
b \\ 0& v& 2vb\\ 0& 0 &\o^{\frac12} \end{array}\right)\ ,
 \ee
where  $\bar b\equiv(\eta~b)^T=b^T\eta$.
The $22\times 22$ sub-matrix $v$ depends
only on the $57$ moduli of $K3$ 
(without the volume) and 
determines the inverse of the matrix 
$M^{ij} = M^i\, \! _k \eta^{kj}$  
introduced in (\ref{omegadual})
\be
M^{-1}=v^T v\ ,\qquad v\eta v^T=\eta\ .
\ee  
The matrices ${\cal M}$ and ${\cal V}$ 
satisfy
\be
{\cal V}{\cal L}{\cal V}^T={\cal L}\ ,\qquad
 {\cal M}{\cal L}{\cal M}^T={\cal L}\ ,\qquad
{\cal M}^T={\cal M}\ ,
\ee
where ${\cal L}$ is the  $O(4,20)$ metric
\be
{\cal L} =\left(\begin{array}{ccc}0 &0& 1\\ 0& \eta& 0\\ 1& 0& 
0\end{array}\right)\ .
\ee

The action \eqn{action} is invariant under
global  $O(4,20)$ transformations acting according to
\bea
&& {\cal M}\to U {\cal M}~ U^T\ ,\qquad {\cal A}^a\to U^a_{~b}~{\cal A}^b\ ,
\qquad m^a\to U^a_{~b}~m^b\ ,
\br 
&&\phi\to\phi\ ,\qquad
B\to B\ ,\qquad g_{\m\n}\to g_{\m\n}\ ,
\lll{odd}
\eea
where $U\in O(4,20)$. These symmetry transformations except the
transformations of the mass parameters are precisely the
T-duality transformations of the standard (massless) type IIA theory on $K3$.
Indeed, in the limit $m^a\to 0$ the action 
(\ref{action}) reduces to the
action of massless type IIA on $K3$ \cite{sen}. 
In the massive case the symmetry 
can be maintained if the 24 mass parameters $m^a$ transform
in the vector representation of $O(4,20)$. This transformation of 
masses  means that under the action 
of the duality group a massive IIA compactified on $K3$ with
fluxes transforms into another massive IIA with a different set
of background fluxes as determined by the symmetry. We emphasize
 that the action \eqn{action} represents a unification of a wide
class of massive supergravities related to each other via the action of
$O(4,20)$ symmetry. Any particular choice of the mass vector $m^a$
represents a
different massive theory. For example, if we set $m^a=(0,\ldots,0,m)$ in
the
action \eqn{action} the theory represents a pure massive IIA compactified
on $K3$ without any fluxes. Similarly a different choice
$m^a=(\tilde m,0,\ldots,0)$ in the above action represents a
six-dimensional
massive theory obtained through generalized reduction of type IIA on $K3$
with 4-form flux. We will show explicitely in the next sections 
that these two theories with single mass parameters are related via
an element of the duality symmetry \eqn{odd}.

Apart from ordinary gauge invariance
${\cal A}_1^a \to {\cal A}_1^a + d\Lambda_0^a$
the field strengths given in  \eqn{deff}, the Bianchi
identities $d{\cal F}^a_2=2m^a ~H_3,~dH_3=0$ as well as the 
action (\ref{action}) 
exhibit  a Stueckelberg type gauge symmetry
of the form
\be
\delta B_2= d\Lambda_1\ ,\qquad \delta {\cal A}^a_1=-2m^a\Lambda_1 \ ,
\ee
where $\Lambda_1$ is a one-form.
This gauge invariance can be used to absorb one of the 24
gauge fields into $B_2$ and render it massive. 
Such a gauge fixed version of the theory breaks the 
$O(4,20)$ symmetry spontaneously
since in a given vacuum only one of the $A$-fields can be absorbed. 
Nevertheless, at the level of
the action  the $O(4,20)$ symmetry is manifest and will play an important
role in determining new vacuum configurations.

\section{Domain Wall solutions}

Generally, massive supergravities admit domain-wall solutions
which preserve half of the supersymmetries. So we also
expect this to be
the case for the six-dimensional
massive theory in \eqn{action}. 
It is known that the  ten-dimensional massive IIA theory has a D-8-brane
(domain-wall) solution which preserves 16 supercharges \cite{berg}. 
In the string frame  metric it is given by
\bea
&&d\hat s^2= H^{-1/2} (-dt^2 + \sum_{i=1}^8dx_i^2) +H^{1/2}dy^2,
\br &&2\hat\f=-{5\ov2}\ln H,
\lll{wall}
\eea
where 
$H=1+{2m}|y-y_0|$ is a harmonic function of the transverse
coordinate
$y$ and  all other fields have vanishing backgrounds.
This solution can be compactified by 
wrapping four of the world-volume directions  on $K3$.  In other
words one can also write a D-8-brane solution  with four of
its brane directions being along $K3$
\bea \label{d8k3}
&&d\hat s^2= H^{-1/2} (-dt^2 + \sum_{i=1}^4dx_i^2 +
ds^2_{K3}) +H^{1/2}dy^2\ ,
\br &&2\hat\f=-{5\ov2}\ln H\ ,
\qquad H=1+{2m}|y-y_0|\ .
\eea
That this is indeed a solution of the equations of motion 
has been shown in \cite{BP}, where it is argued that one can 
replace the spatial part of the D-8-brane's worldvolume 
by any manifold of the form $\mathbb{R}_{8-d} \times K_d$
with a Ricci-flat $K_d$. It is further shown that for
$K_d = K3$ the solution preserves 8 supercharges.\footnote{As anticipated 
in section 2 the solution \eqn{d8k3} is a warped product of a
six-dimensional domain wall and a $K3$.} 
The corresponding  six-dimensional domain-wall solution 
can be written as
\bea\label{sixs}
&&ds_6^2= H^{-1/2} (-dt^2 + \sum_{i=1}^4dx_i^2) +H^{1/2}dy^2\ ,
\br &&2\f= 2\hat\f- \ln \o=-{3\ov2} \ln H \ ,
\br &&m^a=(0,\ldots,0,~m)\ , 
\br &&{\cal M}^{-1}={\rm diag}(H,1,\ldots ,1, H^{-1})\ , 
\eea
while all other six-dimensional background values
vanish. The breathing mode $\omega$ defined in (\ref{volume}) 
is given by $H^{-1}$ in this example. Furthermore in 
(\ref{sixs}) we have chosen a special 
point in the moduli space of $K3$ where
(\ref{omegadual}) reads
\be \label{omegadual2}
^\ast\Omega_2^i = \Omega_2^{23-i}\ , \quad
^\ast \Omega_2^{23-i} = \Omega_2^i\ , \quad  {\rm for}~ i=1,2,3 \ ; \qquad 
^\ast \Omega_2^j = \Omega_2^j\ , \quad {\rm for}~j=4,\ldots,19\ .  
\ee
The solution (\ref{sixs}) still has $8$ unbroken supersymmetries. 

Now, by applying an $O(4,20)$ duality
transformation \eqn{odd} on the background in \eqn{sixs}
new solutions with non-trivial R-R fluxes can be generated.
Let us first consider the special case
where the matrix $U$ is taken to be
\be\label{U3}
U=\left(\ba{ccc} 0&0&1\\ 0&I_{22}&0 \\ 1&0&0 \ea\right)\ .
\ee
Inserting $U$ and the configuration \eqn{sixs} in \eqn{odd} we
get
$\o\to \o'=1/\o\ $ and 
\be
m^a=\left(\ba{c} 0\\ . \\ . \\ 0\\ m\ea\right) \to m'^a=
\left(\ba{c} m\\ 0 \\ . \\ .\\ 0\ea\right)\ ,\qquad
{\cal{M}}^{-1}=\left(\ba{ccc}H&0&0\\ 0&I_{22}&0\\ 0&0& H^{-1}\ea\right)\to 
{{\cal M}'}^{-1}=\left(\ba{ccc}H^{-1}&0&0\\ 0&I_{22}&0\\ 0&0& H\ea\right)\
,
\lll{dual1}
\ee
while the six-dimensional metric and the dilaton  remain the
same. The
transformed mass vector $m'^a$ implies that the new
configuration is a solution of a massless IIA
compactified on $K3$   
with an equivalent amount of four-form flux turned on along $K3$.
When \eqn{dual1} is lifted to ten dimensions we get the following new
configuration
\bea\label{news}
&&d \hat s'^2= H^{-1/2} (-dt^2 +
\sum_{i=1}^4dx_i^2)  +H^{1/2}(dy^2 + ds_{K3}^2 )\ ,
\br &&2\hat\f'=-{1\ov2} \ln  H\ ,\qquad \hat F'_4=-{m} \Omega_4\ .
\eea
Since this solution is obtained by the duality
transformation \eqn{odd} the number of preserved
supercharges is unchanged. It can also be checked explicitly
that \eqn{news} leaves $8$ supercharges unbroken.
Since under the duality transformation \eqn{dual1} $\o\to1/\o$, which
amounts to making T-duality along all the four directions of $K3$,
the background  \eqn{news} represents a stack of D-4-branes 
filling the $K3$ with
non-trivial 4-form flux along $K3$. This
configuration can be further lifted to eleven dimensions as we will
see in the next section. 

By applying an $O(4,20)$ T-duality transformation
we transformed a solution of massive type IIA 
to a solution of massless type IIA
with non-trivial four-form flux.
Thus, the  $O(4,20)$ duality
interpolates between vacua of massive IIA  
and massless IIA. In spirit this is 
similar to the situation encountered 
in the case of massive type II duality in
$D=9$ \cite{berg}. 

Further solutions in $D=6,10$ can be generated 
by using other elements of the duality group 
which mix the mass $m$ with the fluxes $m^i$
of the 2-cycles. Let us
consider the  case
\be
U=\left(\ba{ccccc} 0&1&0&0&0\\ 1&0&0&0&0\\
0&0&I_{20}&0&0\\ 0&0&0&0&1 \\ 0&0&0&1&0 \ea\right)
\ ,
\lll{21a}\ee
which mixes $m$ and $m_{22}$. Note that the duality element in \eqn{21a}
preserves the $O(4,20)$ metric as can be checked using the explicit form 
for $\eta$ given in (\ref{eta}). 
When we apply \eqn{21a} on the configuration
\eqn{sixs}, using \eqn{odd} we
find
$\o\to 1$ and
\be
m^a=\left(\ba{c} 0\\ . \\ . \\ 0\\ m\ea\right) \to
\left(\ba{c} 0\\ . \\ . \\ m\\ 0\ea\right)\ ,
\qquad
{\cal{M}}^{-1}=
\left(\ba{ccc}H&0&0\\ 0&I_{22}&0\\ 0&0& H^{-1}\ea\right)\to 
{\cal{M}}'^{-1}=
\left(\ba{ccccc}1&0&0&0&0\\ 0&H&0&0&0\\ 0&0&I_{20}&0&0\\ 0&0&0&H^{-1}&0\\ 
0&0&0&0&1\ea\right)\ ,
\lll{dual}
\ee
while all other six-dimensional fields remaining unchanged. Again from 
the analysis of the new mass-vector it can be seen that this new
six-dimensional configuration corresponds to massless IIA
compactified on $K3$ but now with a 
two-form flux along a 2-cycle of $K3$. When lifted to  ten dimensions we
have a new solution of massless IIA theory as
\bea\label{cdf}
&&d \hat s'^2= H^{-1/2} (-dt^2 +
\sum_{i=1}^4dx_i^2)  + ds'^2_{K3} +H^{1/2}dy^2\ , \br 
&&2\hat\f'=-{3\ov2} \ln  H\ , \qquad
\hat F'_2= {m} \Omega_2^{(22)}\ ,
\eea
where  $ds'^2_{K3}$ is the deformed metric on $K3$ such that the
deformation corresponds to the nontrivial moduli matrix 
${\cal{M}}'^{-1}$ in \eqn{dual}.
$\Omega_2^{(22)}$ is a  2-cycle on $K3$. 
This background configuration \eqn{cdf}  preserves the same amount of
supersymmetries as the one in \eqn{sixs}.

One can go through a similar analysis for solutions which depend
on more than one mass or flux parameter.
Let us consider the following solution of 
the massive IIA theory in \eqn{1a} 
\bea
&&d \hat s^2= H^{-1/2}H'^{-1/2} (-dt^2 +
\sum_{i=1}^4dx_i^2)+H^{1/2}H'^{-1/2}ds^2_{K3} 
+H^{1/2}H'^{1/2}dy^2\ ,
\br &&2\hat\f=-{1\ov2} \ln  (H H'^5)\ ,\qquad
\hat F_4= -\tilde m ~\Omega_4\ ,\qquad \hat F_2=0=\hat B_2\ ,
\br
&& H'=1+2m|y|\ ,\qquad H=1+2\tilde m|y|\ .
\eea
For the special case when $\tilde m=m,~H'=H$ it reduces to
\bea
&&d \hat s^2= H^{-1}(-dt^2 + \sum_{i=1}^4dx_i^2)  +H~dy^2 +ds^2_{K3}\ ,
\br &&2\hat\f=-3 \ln  H\ ,\qquad
\hat F_4= - \tilde m ~\Omega_4\ ,\qquad \hat F_2=0=\hat B_2\ ,
\eea
which preserves $8$ supercharges as can be checked explicitly. This can be
compactified to six dimensions
and using the  $U$-matrix of \eqn{21a}
it can be transformed as 
\bea
&& m^a=\left(\ba{c} m\\0\\ . \\ . \\ 0\\ m\ea\right) \to
\left(\ba{c} 0\\m_1=m\\0\\ . \\ .\\0 \\ m_{22}=m\\ 0\ea\right)\ ,\qquad 
{\cal M}^{-1}=
\left(\ba{ccc}1&0&0\\ 0&I_{22}&0\\ 0&0& 1\ea\right)\to 
\left(\ba{ccc}1&0&0\\ 0&I_{22}&0\\ 0&0&1\ea\right).
\lll{dual11}
\eea
Lifting this solution back to ten dimensions we get
\bea
&&d \hat s^2= H^{-1}(-dt^2 + \sum_{i=1}^4dx_i^2)  +H~dy^2
+ds^2_{K3}\ ,
\br &&2\hat\f=-3 \ln  H\ ,\qquad
\hat F_4= 0\ ,\qquad \hat F_2=m_1\Omega_2^{(1)}+m_{22}\Omega_2^{(22)}.
\eea
$\hat F_2$ is self-dual since $m_1=m_{22}=m$. 
This is the solution obtained in \cite{lavpop} for massless type IIA.

\section{A Lift to M-theory}
It is conjectured that the strong coupling 
limit of  type IIA is governed by  M-theory \cite{witten} whose low energy
limit is believed to be 11-dimensional supergravity \cite{cjs}. In
fact one
can obtain massless type IIA supergravity as an 
$S^1$ compactification of 11-dimensional supergravity and in this way all
$p$-brane solutions of 
massless type IIA can be lifted to eleven dimensions. 
However, the  8-brane solution of massive IIA
given in \eqn{wall}
cannot  be lifted easily to 11-dimensions since
that would require 
a massive version of 11-dimensional
supergravity 
\cite{blo,bs} which does not exist \cite{deser,deser2}. 
In ref.\ \cite{hull} a general relation between
massive IIA theory and  M- and F-theory has been proposed.
We will see here that taking the detour of
compactifying massive IIA on $K3$  provides another
possibility to relate the 8-brane solution
to M-theory.\footnote{It would be interesting
to understand the connection with
ref.\ \cite{hull} in more detail.}
In order to show this in slightly more detail 
let us write down
the map among the fields of massive and  massless
type IIA on $K3$.
As discussed above,
 the six-dimensional theory in
\eqn{action} with the mass vector $m^a=(\tilde m,0, \ldots,0)$ represents
an ordinary type IIA on $K3$ with RR-4-flux. On the other hand
the theory with mass vector $m^a=(0,\ldots,0,m)$ represents
massive IIA on $K3$ without any RR-flux. These two massive
six-dimensional theories are
related by the duality element \eqn{U3}.
In section~3,
eqs.\ \eqn{sixs}--\eqn{news} we displayed the T-duality between 
a domain wall solution of massive IIA and a stack of solitonic D-4-branes of 
massless type IIA. In fact one can not only map
the solutions onto each other
but the entire actions.
Under the transformation 
given in \eqn{U3}
massive IIA  with no fluxes transforms into massless IIA with 4-form flux
as 
\bea
&& \o^{-1}\to
\o+4b^TM^{-1}b+{4(b^2)^2\over\o},\br
&&-{2\bar b\over\o}\to{4b^2\bar b\over\o}+2b^TM^{-1},~~M^{-1}\to
M^{-1},~~~{2b^2\over\o}\to{2b^2\over\o},\br
&&d\tilde A\to dA+{2m}B,~~~dA+{2m}B\to d\tilde
A,~~~dC^i\to dC^i,\br
&&\phi\to\phi,~~~B\to B,~~~~g_{\m\n}\to g_{\m\n}.
\lll{mapp}
\eea
Under this map 
the mass parameter $m$ of massive type IIA
is mapped to the four-form flux $\tilde m$ on $K3$ of massless IIA and
vice versa. 
 
Since massless type IIA on 
$K3$ is equivalent to M-theory on 
$S^1\times K3$, all solutions of
massive IIA can be lifted to eleven dimensions by first mapping
them to solutions of the massless theory using  the map 
\eqn{mapp}.
The solution given in \eqn{news} 
corresponds to the following eleven dimensional solution 
\bea\label{M5}
ds_{(11)}^2&=&e^{4\hat\f\over3}(dx_{11}+
\hat A^Mdx_M)^2+e^{-{2\hat\f\over3}}ds_{(10)}^2  \br
&=&H^{-{1\ov3}} (-dt^2 +
\sum_{i=1}^4dx_i^2+ dx_{11}^2 ) +H^{2\over3}(dy^2+ds_{K3}^2)\ ,\br
G_4&=&-m~ \Omega_4\ ,\qquad H=1+2m|y|\ .
\eea
where $G_4$ is the 4-form field strength of
eleven-dimensional supergravity.
The solution \eqn{M5} is a stack of M5-branes which couple magnetically 
to the 4-form field strength.\footnote{ 
It is possible to establish a similar map as in \eqn{mapp}
between the backgrounds of
massive IIA compactified on $K3$ (with $ m^i\ne 0$)
and the backgrounds
of massless IIA on $K3$ (with $ m^i\ne 0$). However, it is not clear whether 
those can be lifted to 11 dimensions since this would 
require a globally defined $\hat A_1$.}
 
Thus it seems that the conjectured duality 
between type IIA compactified on
$K3$ and M-theory  on $K3\times S^1$
extends to  the massive case.
It is shown in
ref.\ \cite{lupop}
that for M-theory on $K3\times S^1$ the compactifications on $K3$ and
$S^1$ commute even in the presence of 4-form flux
on $K3$.
Here we have seen that massless
type IIA compactified on $K3$  
with `4-form flux along $K3$' is T-dual to
massive IIA  compactified on $K3$ without flux. 
 Thus we are led to conjecture that
M-theory  on $K3\times S^1$ with 
`4-form flux on $K3$' is dual to
massive type IIA compactified 
on a $K3$ without flux. In this duality the mass of Romans theory is
mapped to the 4-form flux of M-theory along $K3$.

\vskip1cm

\begin{tabular}{rcl}
~&~~~ & \fbox{\large\rm M-theory~ on~ $S^1\times K3$ $\&$ 
4-Flux}\\
~&~~~& ~~~~~~~~~$\updownarrow$ ~\\    
\fbox{\large\rm massive\ IIA\ on~ $K3$} & $\stackrel{(\o\leftrightarrow 
{1\over \o})}
{\longleftrightarrow}$
&\fbox{\large\rm massless\ IIA~ on~ $K3$ $\&$ 4-Flux}
\end{tabular}
\vskip2cm

\section {Conclusion}
In this paper we derived the action for 
$K3$ compactifications of 
ten-dimensional massive type IIA theory 
with all Ramond-Ramond background fluxes 
turned on. We found that the resulting six-dimensional theory is a new
massive (gauged) supergravity with an 
action having manifest $O(4,20)\over O(4)\times O(20)$ duality symmetry
provided the mass (flux) parameters transform accordingly.
>From this  we learn that the perturbative T-duality survives even at
the massive level when appropriate masses and fluxes are switched
on.  

We have seen in this paper that the massive six-dimensional theory
of \eqn{action}
interpolates between ten-dimensional
massive and massless type IIA theories. The wrapped 
D-8-brane solution of massive
type IIA turns out to be T-dual
to a supersymmetric solution of massless
IIA theory on $K3$ with four-form flux.
The relationship between massless and massive IIA on $K3$ also 
suggests a new 11-dimensional interpretation of massive IIA theory.

As it is known, in addition to the perturbative T-duality
there also is a non-perturbative
S-duality between massless type IIA on $K3$
and massless heterotic string on $T^4$ \cite{harvey,DLM,sen}.
Both theories have the same 
perturbative T-duality group 
$O(4,20)\over O(4)\times O(20)$
and their respective dilatons are related by  
$\phi_{Het}\to -\phi_{IIA}$. 
Let us recall that there also exists an 
$O(d,d+n)$  symmetric
massive compactification of the heterotic string 
on $T^d$ \cite{km}.
However, the non-perturbative S-duality 
seems no longer
to be valid in the massive theories.
The major difference is that in the type IIA
theory \eqn{action} it is the tensor
field B which becomes massive after eating 
the vector fields while in the 
heterotic theory  vector fields 
become massive after eating
some scalars \cite{km}. This is similar to the situation encountered in 
the duality  
between massive M-theory on $K3$ and heterotic theory on $T^3$
\cite{lavpop}.    
Furthermore there is a runaway dilaton potential in both theories
driving the dilaton to weak coupling. As a consequence strong-weak
S-duality can no longer hold. 

Finally, it is interesting to consider a further 
reduction of the massive
six-dimensional theory \eqn{action} on $S^1$ and explore some duality
relationship with type IIB on $K3\times S^1$ with fluxes. It is clear from
the action \eqn{action} that there are no axions in this theory so the
further reduction on $S^1$ will not produce any new mass parameter. On
the other hand the standard reduction of type IIB on $K3$ produces
exactly  
24 axions in six dimensions which upon further generalized
compactification on $S^1$
will generate 24 mass parameters in five dimensions. Through a duality
relation one should be
able to relate the 24 mass parameters of massive IIA on $K3\times S^1$ to
those of type IIB on $K3\times S^1$. This should be related to the
duality of massive type
IIA and type IIB in $D=9$ \cite{berg}.

\pagebreak
 
\noindent
{\bf ACKNOWLEDGMENTS}

\noindent
The work of M.H.\  is supported by 
the DFG (the German Science Foundation)
and the DAAD (the German Academic Exchange Service). 
The work of J.L.\  is supported by 
GIF (the German--Israeli 
Foundation for Scientific Research),
the DFG and the DAAD.
The work of H.S.\
is supported by
AvH (the Alexander von Humboldt foundation).

\end{document}